
%
\input harvmac
%
%
%
\def\RF#1#2{\if*#1\ref#1{#2.}\else#1\fi}
\def\NRF#1#2{\if*#1\nref#1{#2.}\fi}
\def\refdef#1#2#3{\def#1{*}\def#2{#3}}
\def\rdef#1#2#3#4#5{\refdef#1#2{#3, {\it #4}, #5}}
%
%
\def\ts{\hskip .16667em\relax}

\def\CMP{{\it Comm.\ts Math.\ts Phys.\ts}}

\def\IJMP{{\it Int.\ts J.\ts Mod.\ts Phys.\ts}}

\def\NP{{\it Nucl.\ts Phys.\ts}}
\def\PL{{\it Phys.\ts Lett.\ts}}

\def\Tahoe{Proceedings of the XVIII International Conference on
 Differential Geometric Methods in Theoretical Physics: Physics
 and Geometry, Lake Tahoe, USA 2-8 July 1989}
\def\Tahoe{Proceedings of the NATO
 Conference on Differential Geometric Methods in Theoretical
 Physics, Lake Tahoe, USA 2-8 July 1989 (Plenum 1990)}
\def\Zm{Zamolodchikov}
\def\AZm{A.\ts B.\ts \Zm}
\def\AlZm{Al.\ts B.\ts \Zm}
\def\me{P.\ts E.\ts Dorey}
\def\dur{H.\ts W.\ts Braden, E.\ts Corrigan, \me\ and R.\ts Sasaki}
%
%
\rdef\rAFZa\AFZa{A.\ts E.\ts Arinshtein, V.\ts A.\ts Fateev and \AZm}
{Qunatum S-matrix of the 1+1 dimensional Toda chain}
{\PL {\bf B87} (1979) 389}
\refdef\rBCa\BCa{R. Brunskill and A. Clifton-Taylor, {\it English Brickwork}
 (Hyperion 1977)}
\rdef\rBCDSb\BCDSb{\dur}
{Aspects of perturbed conformal field theory, affine Toda field
theory and exact S-matrices}
{\Tahoe}
\rdef\rBCDSc\BCDSc{\dur}
{Affine Toda field theory and exact S-matrices}
{\NP {\bf B338} (1990) 689}
\rdef\rBCDSe\BCDSe{\dur}
{Multiple poles and other features of affine Toda field theory}
{\NP {\bf B356} (1991) 469}
\refdef\rBCDSf\BCDSf{\dur, \NP {\bf B338} (1990) 689;
 \NP {\bf B356} (1991) 469}
\refdef\rBSa\BSa{H.\ts W.\ts Braden and R.\ts Sasaki,
{\it The S-matrix coupling dependence for a, d and e affine Toda
field theory}, \PL {\bf B255} (1991) 343;\quad
{\it Affine Toda perturbation theory}, \NP {\bf B379} (1992) 377}
\refdef\rCe\Ce{E. Corrigan, private communication}
\rdef\rCMa\CMa{P.\ts Christe and G.\ts Mussardo}
{Integrable systems away from criticality: the Toda field theory and S
matrix of the tricritical Ising model}
{\NP {\bf B330} (1990) 465}
\rdef\rCMb\CMb{P.\ts Christe and G.\ts Mussardo}
{Elastic S-matrices in (1+1) dimensions and Toda field theories}
{\IJMP {\bf A5} (1990) 4581}
\refdef\rCSa\CSa{E. Corrigan and R. Sasaki, private communication}
\refdef\rCTa\CTa{S.\ts Coleman and H.\ts Thun, \CMP {\bf 61}
 (1978) 31}
\refdef\rDb\Db{\me, Durham PhD thesis, unpublished}
\rdef\rDc\Dc\me
 {Root systems and purely elastic S-matrices, I \& II}
 {\NP {\bf B358} (1991) 654; \NP {\bf B374} (1992) 741}
\rdef\rDd\Dd\me
 {Root systems and purely elastic S-matrices II}
 {\NP {\bf B374} (1992) 741}
\refdef\rDf\Df{\me, \NP {\bf B358} (1991) 654; \NP {\bf B374} (1992) 741}
\rdef\rDg\Dg\me
 {Partition Functions, Intertwiners and the Coxeter Element}
 {preprint SPhT/92-053, hep-th/9205040; \IJMP {\bf A}, in press}
\rdef\rDh\Dh\me
 {Hidden geometrical structures in integrable models}
 {preprint NI\ts 92018, hep-th/9212143, CERN-TH.6772/93, to appear in
 the proceedings of the conference ``Integrable Quantum Field Theories'',
 Como, Italy, 13-19 September 1992}
\rdef\rDDa\DDa{C.\ts Destri and H.\ts J.\ts de Vega}
{The exact S-matrix of the affine $E_8$ Toda field theory}
{{\it Phys. Lett.} {\bf B233} (1989) 336}
\refdef\rDGPZa\DGPZa{G.\ts Delius, M.\ts Grisaru, S.\ts Penati and D.\ts
Zanon, \PL {\bf B256} (1991) 164\semi \NP {\bf B359} (1991) 125}
\rdef\rDGZa\DGZa{G.\ts W.\ts Delius, M.\ts T.\ts Grisaru and D.\ts Zanon}
 {Exact S-matrices for non simply-laced affine Toda theories}
 {\NP {\bf B282} (1992) 365}
\rdef\rFb\Fb{M.\ts D.\ts Freeman}
{On the mass spectrum of affine Toda field theory}
{\PL {\bf B261} (1991) 57}
\rdef\rFKMa\FKMa{P.\ts G.\ts O.\ts Freund, T.\ts Klassen and E.\ts Melzer}
{S-matrices for perturbations of certain conformal field theories}
{\PL {\bf B229} (1989) 243}
\rdef\rFLOa\FLOa{A.\ts Fring, H.\ts C.\ts Liao and D.\ts I.\ts Olive}
{The mass spectrum and coupling in affine Toda theories}
{\PL {\bf B266} (1991) 82}
\refdef\rKe\Ke{V.\ts Kac, {\it Infinite dimensional Lie algebras}
(Cambridge University Press 1990)}
\rdef\rKWb\KWb{H.\ts G.\ts Kausch and G.\ts M.\ts T.\ts Watts}
{Duality in quantum Toda theory and $W$-algebras}
{\NP {\bf B386} (1992) 166}
\rdef\rMOPa\MOPa{A. V. Mikhailov, M. A. Olshanetsky and A. M. Perelomov}
{Two-dimensional generalised Toda lattice}
{\CMP {\bf 79} (1981) 473}
\rdef\rOTa\OTa{D.\ts I.\ts Olive and N.\ts Turok}
{The symmetries of Dynkin diagrams and the reduction of Toda field
equations}
{\NP {\bf B215} (1983) 470}
\rdef\rSf\Sf{F.\ts Smirnov}
{Reductions of the sine-Gordon model as a perturbation of minimal
models of conformal field theory}
{\NP {\bf B337} (1990) 156}
\refdef\rWa\Wa{G. Wilson, {\it Ergod. Th. and Dynam. Sys.} {\bf 1} (1981) 361}
\rdef\rWWa\WWa{G.\ts M.\ts T.\ts Watts and R.\ts A.\ts Weston}
{$G_2^{(1)}$ affine Toda field theory. A numerical test of exact
S-matrix results}
{\PL {\bf B289} (1992) 61}
\rdef\rZa\Za{\AZm}
{Integrable Field Theory from Conformal Field Theory}
{Proceedings of the Taniguchi Symposium, Kyoto (1988)}
\rdef\rZb\Zb{\AZm}
{Integrals of motion and S-matrix of the (scaled) $T{=}T_c$ Ising model with
magnetic field}
{\IJMP {\bf A4} (1989) 4235}
\refdef\rZd\Zd{\AZm, {\it Int. J. Mod. Phys.} {\bf A3} (1988) 743}
\refdef\rZe\Ze{\AZm, {\it Sov. Sci. Rev., Physics}, {\bf v.2} (1980)}
\refdef\rZz\Zz{\Za\semi\Zb}
\rdef\rZZa\ZZa{\AZm\  and \AlZm}
{Factorized S-matrices in two dimensions as the exact solutions of certain
relativistic quantum field theory models}
{{\it Ann. Phys.} {\bf 120} (1979) 253}
\rdef\rCDSa\CDSa{E.\ts Corrigan, P.\ts E.\ts Dorey and R.\ts Sasaki}
{On a generalised bootstrap principle}
{preprint DTP-93/19, YITP/U-93-09, CERN-TH.6870/93; hep-th/9304065}
%
%
\def\aa{a_{2n-1}^{(2)}}
\def\aae{a_{2n}^{(2)}}
\def\bb{b_{n}^{(1)}}
\def\cc{c_{n}^{(1)}}
\def\dd{d_{n+1}^{(2)}}
\def\ddd{d_4^{(3)}}
\def\ee{e_6^{(2)}}
\def\ff{f_4^{(1)}}
\def\gg{g_2^{(1)}}
\def\gk{g^{(k)}}
\def\hv{h^{\vee}}
\def\hk{h^{(k)}}
\def\hkv{\hk{}^{\vee}}
\def\rsum{\sum_{i=0}^n}
\def\({\left(}
\def\){\right)}
\def\Bgen#1{B^{[#1]}}
\def\Bgk{\Bgen{\gk}}
\def\bar{\overline}

\def\forget#1{}
%
%
   
%

%

%

%

%
\def\a{\alpha}

%
%
\def\frac#1/#2{\leavevmode\kern.1em
\raise.5ex\hbox{$\scriptstyle #1$}\kern-.1em
/\kern-.15em\lower.25ex\hbox{$\scriptstyle #2$}}

\def\th{\ifmmode{^{\rm th}}\else{$^{\rm th}$}}

%
%

\def\\{\relax\ifmmode\backslash\else$\backslash$\fi}
%
%
\def\usbl#1{\hbox{{\bf (}$#1${\bf )}}}
\def\ubl#1{\{#1\}}

\def\<{\langle}
\def\>{\rangle}
\def\bl#1{\<#1\>}

\def\ibl#1#2#3{\vphantom{\ubl{#2}}_{\lower 2pt\hbox%
{$\scriptstyle #1$}}\ubl{#2}_{\lower 2pt\hbox{$\scriptstyle #3$}}}
\def\tc{\times{\rm crossing}}
%
%
\def\clap#1{\hbox to 0pt{\hss #1\hss}}
\def\ulabel#1{\raise3ex\clap{$\scriptstyle #1$}}
\def\dlabel#1{\lower2ex\clap{$\scriptstyle #1$}}
%
%
\def\va{\noalign{\vskip .15in}}
\def\vb{\noalign{\vskip .06in}}
\def\hrle{\noalign{\hrule}}
\def\tableone{\vbox{\tabskip=0pt\offinterlineskip
\halign{~~##\hfil&##\hfil&##\hfil&##\hfil&##\hfil&##\hfil\cr
\hrle
\vb
$\gk$ & $k$ & $\hk$ & $\hkv$& $h$ & $h^{\vee}$ \cr
\vb
\hrle
\va
$\bb$&$1$&$2n$&$4n\!-\!2$&$2n$&$2n\!-\!1$\cr
\vb
$\aa$&$2$&$4n\!-\!2$&$2n$&$2n\!-\!1$&$2n$\cr
\va
$\cc$&$1$&$2n$&$2n\!+\!2$&$2n$&$n\!+\!1$\cr
\vb
$\dd$&$2$&$2n\!+\!2$&$2n$&$n\!+\!1$&$2n$\cr
\va
$\ff$&$1$&$12$&$18$&$12$&$9$\cr
\vb
$\ee$&$2$&$18$&$12$&$9$&$12$\cr
\va
$\gg$&$1$&$6$&$12$&$6$&$4$\cr
\vb
$\ddd$&$3$&$12$&$6$&$4$&$6$\cr
\va
$\aae~~~~~~~~~$&$2~~~~~~~$&$4n\!+\!2~~~~~~$%
&$4n\!+\!2~~~~~~$&$2n\!+\!1~~~~~~$&$2n\!+\!1~~$\cr
\va
\hrle
\va
&\rlap{\quad Table 1 : Lie algebra data}&&&&\cr
}}}
%
%
%
%
%
%
\ifx\answ\bigans%
\else\def\tfontsize{scaled\magstep3}
 \tfontsize  \tfontsize
 \tfontsize  \tfontsize
 \tfontsize  \tfontsize
 \tfontsize  \tfontsize
 \tfontsize  \tfontsize
\hstitle=\hsbody\fullhsize=10.4truein\hsize=\hsbody
\output={\almostshipout%
{\leftline{\vbox{\pagebody\makefootline}}}\advancepageno}
\fi
%
%
%
\Title{\vbox{\baselineskip12pt\hbox{CERN-TH.6873/93}%
\hbox{hep-th/9304149}}}%
{\vbox{\centerline{A remark on the coupling-dependence}\vskip2pt
       \centerline{in affine Toda field theories}}}
\centerline{Patrick Dorey}
\smallskip
{\footnotefont
\centerline{CERN TH, 1211 Geneva 23, Switzerland}
\centerline{\tt dorey@surya11.cern.ch}
\vskip .3in
\baselineskip=12.6pt plus 2pt minus 1pt
The affine Toda field theories based on the non simply-laced Lie
algebras are discussed. By rewriting the S-matrix formulae found by
Delius {\it et al}, a universal form for the coupling-constant dependence
of these models is obtained, and related to various general properties
of the classical couplings. This is illustrated via the S-matrix
associated with the dual pair of algebras $\ff$ and~$\ee$.
}

\bigskip
\Date{\vbox{\baselineskip12pt\hbox{CERN-TH.6873/93}%
\hbox{April 1993}}}
\baselineskip=14pt plus 2pt minus 1pt

\newsec{Introduction}
For each affine Lie algebra $\gk$ of rank $n$, a classically integrable
field theory in two dimensions can be defined via the Lagrangian
\eqn\todlag{\eqalign{{\cal L}&={1\over 2} (\partial\phi)^2-V(\phi)\,;\cr
V(\phi)&={m^2\over\beta^2}\rsum n_ie^{\beta\a_i\cdot\phi}.\cr}}
The $n$-dimensional vectors $\{\a_0,\dots\a_n\}$ together make up an
(extended) set of simple roots, their inner products being encoded in
the affine Dynkin diagram of $\gk$. The number $m$ sets the overall
mass scale, while $\beta$ is the coupling constant, taken to be
real below. The relative values of the $n_i$ can be changed by a shift
in the field $\phi$; it is convenient to impose $\sum n_i\a_i{=}0$ -- so
that $\phi{=}0$ minimises the potential -- and conventional to choose
$n_0{=}1$. The $n_i$ then agree with the labelling of the affine
diagrams in the book by Kac\ts\RF\rKe\Ke.
The resulting model is known as the affine Toda field theory based on
$\gk$, and its integrability is reflected in the presence of infinitely
many conserved quantities, at spins given by the exponents of the
affine algebra. If these conservation laws survive quantisation, then
\todlag\ should also define a {\it quantum} integrable field theory,
and its scattering amplitudes should factorise into products of
two-particle S-matrices. As part of the general programme to
understand two-dimensional quantum field theories, it is rather
natural to ask what these two-particle S-matrix elements might
be\ts\NRF\rAFZa\AFZa\NRF\rCMa\CMa\NRF\rCMb\CMb\NRF\rBCDSc\BCDSc%
\NRF\rBCDSe\BCDSe\NRF\rDDa\DDa\NRF\rDGZa\DGZa\refs{\rAFZa{--}\rDGZa}.

A straightforward way to approach this question starts by expanding
the potential $V(\phi)$ in $\beta,$ as follows:
\eqn\pertexp{V(\phi)={m\sp 2\over \beta\sp 2}\rsum n_i+
{(M^2)^{ab}\over 2}\phi^a\phi^b+
{C^{abc}\over 3!}\phi^a\phi^b\phi^c +\dots}
The leading non-trivial terms to appear are the classical
$({\rm mass})^2$ matrix, and a collection of classical
three-point couplings:
\eqn\massandcoupl{
(M\sp 2)\sp {ab}=m\sp 2\rsum \alpha_i^a\alpha_i^b \quad;\quad
C^{abc}=m^2\beta\rsum \alpha\sb i\sp a\alpha\sb i\sp b\alpha\sb i\sp c\,.}
These objects have a number of remarkable `universal'
properties\ts\refs{\rCMa{--}\rBCDSc}. If
$\gk$ is untwisted (that is,
if $k{=}1$), then the set of classical masses
$(m_1,m_2,\dots m_n)$ forms a right eigenvector of the
Cartan matrix of the corresponding non-affine algebra $G$ (taking
$C_{ab}=2\a_a\cdot\a_b/\a_b^2\,$; $a,b{=}1\dots n$). For the couplings,
there is a simple rule determining when they are non-vanishing,
relying on the action of a Coxeter element in the relevant (finite)
root system\ts\RF\rDc\Dc\ (a twisted Coxeter element if
$k{>}1$\ts\RF\rDh\Dh).  The magnitudes of the
non-vanishing couplings are given by the `area law'
\eqn\area{|C^{abc}|=\lambda^{abc}{2\beta\over\surd\hk} m_am_b\sin
U^c_{ab}\,,}
where the fusing angle $U^c_{ab}$ is defined via the relation
\eqn\fusang{m_c^2=m_a^2+m_b^2+2m_am_b\cos U^c_{ab}\,.}
The factor $\lambda^{abc}$ is included both to take into account the
normalisation of the roots $\a_i$, and to allow for certain adjustments
necessary in the untwisted non simply-laced cases; it will be
reviewed in more detail later.  The so-called `$k^{th}$ Coxeter number'
$\hk$ is defined as follows\ts\rKe\ts: if the usual
Coxeter number of $\gk$ is $h{=}\sum_0^rn_i$, then $\hk{=}k.h$
(table~1 reproduces some of the relevant data from \rKe\ts).
This quantity is relevant to the classical theory in one further way:
all the fusing angles $U^c_{ab}$ turn out to be integer multiples of
$\pi/\hk$. General proofs of these classical results now
exist\ts\RF\rFb{\Fb\semi\FLOa}.
\midinsert
$$\tableone$$
\endinsert

Armed with this information, the quantum theory can be examined. A
combination of perturbation theory, based on \pertexp, and general
(non-perturbative) principles\ts\NRF\rZZa\ZZa\NRF\rZz\Zz\refs{\rZZa,\rZz},
has led to conjectures for the {\it exact} S-matrices for all of the
affine Toda models\ts\NRF\rCDSa\CDSa\refs{\rAFZa{--}\rDGZa,\rCDSa}.
In the cases where $\gk$ is simply-laced, so that all of the $\a_a$
have the same length, the story seems to be particularly
simple\ts\refs{\rAFZa{--}\rDDa}. At least to one loop, the mass ratios
do not renormalise, and so the classical fusing angles can be assumed
to be directly relevant to the quantum theory as well. This allows the
positions of both simple and higher poles in the S-matrix elements to be
predicted, the latter via an
analysis of the on-shell diagrams responsible for Landau singularities.
Add in some zeroes (at coupling-constant dependent positions) to
ensure that the S-matrix tends to the identity as $\beta\rightarrow
0$, demand agreement to lowest order in perturbation theory, and a
surprisingly uniform collection of ans\"atze emerges,
each S-matrix element being
written as a product of elementary building blocks $\ubl{x}$,
\eqna\slbldef
$$\ubl{x}={\usbl{x\!-\!1}\usbl{x\!+\!1}\over
\usbl{x\!-\!1\!+\!B}\usbl{x\!+\!1\!-\!B}}\quad;\quad
\usbl{x}={\sinh\(\frac{\theta}/2+\frac{i\pi x}/{2\hk}\)
\over\sinh\(\frac\theta/2-\frac i\pi x/{2\hk}\)}~.\eqno\slbldef{a,b}$$
It is important to note here that,
essentially on account of the earlier observation that all the fusing
angles are integer multiples of $\pi/\hk$, the parameters $x$ for
these blocks are always integers.
Expressed in terms of such blocks, the $\beta$-dependence of the
S-matrix elements is completely hidden: even inside the blocks
it only appears via the function $B(\beta)$. Tree-level perturbation
theory dictates that $B(\beta)=\beta^2/2\pi+\dots\,$, and a
natural conjecture, now backed up by higher-order perturbative
calculations\ts\RF\rBSa\BSa, has
\eqn\slBdef{B(\beta)=2\beta^2/(\beta^2+4\pi)\,.}
The combination of \slbldef a\ and \slBdef\
implies a strong-weak coupling duality for the simply-laced models, in
that the S-matrix is unchanged on replacing $\beta$ by
$4\pi/\beta\,$: the operation simply sends $B$ to $2{-}B$ and
leaves all the blocks $\ubl{x}$ unchanged.

Excepting the theory based on $\aae$\ts\rCMb, the picture for the non
simply-laced cases is more complicated. The mass ratios do not
remain fixed; rather, one-loop corrections can already be seen to give
them a non-trivial dependence on the coupling $\beta$.
At first, this seemed to be
an insurmountable obstacle to the construction of a consistent
diagonal S-matrix for any of these models. However, more recent
proposals by Delius {\it et al\/} have shown that this is not necessarily
the case\ts\rDGZa. The physical-strip poles no longer
have fixed positions, but move as the coupling
constant varies. The price paid for this extra flexibility is
that some of the expected bootstrap relations\ts\RF\rZz\zz\ are no longer
obeyed, corresponding to certain poles in the S-matrix elements
which are not at their expected positions even after the
renormalisations of the mass ratios have been taken into account.
Exactly in these situations, Delius {\it et al\/} were able
to identify collections of Landau singularities which modify initial
expectations in a delicate way (a similar phenomenon in the sine-Gordon
model between the one- and two- breather thresholds had also been
observed by Smirnov\ts\RF\rSf\Sf). Furthermore, these poles
can also be understood within a pure S-matrix context, by means of a
particular generalisation of the Coleman-Thun
mechanism\ts\RF\rCDSa\CDSa. While subtleties remain, enough supporting
evidence now exists to leave little doubt that the
S-matrices presented in \rDGZa\ are correct.

One especially interesting feature is the way that the
strong-weak coupling duality appears to
extend\ts\NRF\rKWb\KWb\refs{\rDGZa,\rKWb}. On general
grounds, one might expect that a fuller statement of the simply-laced
duality is that the S-matrix should be unchanged under
the simultaneous transformations $\beta\rightarrow 4\pi/\beta\,$;
$\a_i\rightarrow\a_i^{\vee}={2\over\a_i^2}\a_i\,$. Since the
simply-laced roots were implicitly assumed to have had (length)$^2$
two, this duality reduces to the previous version in these cases. It
groups the non simply-laced theories into the pairs indicated in
table~1, corresponding to
mutually dual affine Dynkin diagrams -- with the exception of
$\aae$, whose diagram is self-dual (throughout, the addition of
${}^{\vee}$ to a symbol indicates the corresponding dual object).
For example, the
large-$\beta$ mass spectrum of one member of such a pair should
reproduce the small-$\beta$ mass spectrum of its partner,
a property that has now been checked numerically for the
$g^{(1)}_2/d^{(3)}_4$
pair\thinspace\RF\rWWa\WWa. But beyond this, the
S-matrices should also be pairwise equivalent under the replacement
$\beta\rightarrow 4\pi/\beta$. The fact that the S-matrices
proposed by Delius {\it et al\/} did indeed turn out to have this
property -- it had not been fed in at the start -- provided further
support for their conjectures.

The aims of this note are twofold: on the one hand, to systematise previous
results with a `naturally dual' notation applicable to both simply-
and non simply- laced cases, in the process observing the simple
generalisation of \slBdef; and on the other to illustrate this with
some features of the $\ff$/$\ee$ S-matrix.

\newsec{A general block notation for all affine Toda theories}
In \rDGZa, the coupling dependence of the pole positions
was incorporated in two ways: first, by allowing
$\hk$ to wander away from its `classical' ($k^{th}$ Coxeter
number) value, replacing it in \slbldef b\ by $H(\beta)$,
the `renormalised Coxeter number'; and second, by using parameters
$x$ in the blocks defined by \slbldef a\ that were no longer
always integers, but rather could depend on $\beta$ via $H(\beta)\,,$
and sometimes also via $B(\beta)$. Most, though not in fact quite all,
of the S-matrix elements were then written in terms of such blocks
alone.

The alternative to be advocated here starts by
dropping the idea that $B(\beta)$ should always be given by the
expression \slBdef. (With standard normalisations for the roots, this
is in any case impossible to enforce: cf.~\rWWa\ for the
$g_2^{(1)}$ case.) Instead,
for the theory based on $\gk$, define
$\Bgk(\beta)$ to be that function of $\beta$ which satisfies
\smallskip
\item{(a)} $\Bgk(0)\!=\!0~,~~\Bgk(\infty)\!=\!2\,$;
\item{(b)} The positions of the zeroes and poles in the S-matrix
elements of $\gk$ depend {\it linearly} on $\Bgk\,$.
\smallskip
\noindent Given the S-matrix, this specifies $\Bgk$ uniquely,
and it reproduces \slBdef\ in the simply-laced cases. Strictly
speaking, just a single mobile pole or zero is enough to pin $\Bgk$
down, so part (b) relies on the observation that
all poles and zeroes can be linearised by a
single function. If this seems surprising, it should be no more
so than the fact that a single function $B(\beta)$ sufficed in each of
the simply-laced models; the underlying reason is
that the bootstrap relations, binding the different S-matrix
elements together in a set of overdetermined equations, forces the
coupling-dependence to enter in a coherent way.

An example: in the $\aa$ theory, Delius {\it et al\/} found that
$S_{nn}(\theta)$ had (amongst others) a pole at $2/(2n{-}1{+}B/2)$,
when $B(\beta)$ is given by \slBdef. So, this $B$ fails on count (b).
But
\eqn\atnewBa{{2\over 2n{-}1{+}B/2}={2\over 2n{-}1}+{-B\over
(2n{-}1)(2n{-}1{+}B/2)}~,}
and the second term on the RHS does satisfy (b) above, along with the
first half of (a); multiplying it by $-2n(2n{-}1)$
to give it the desired limiting value of $2$ as
$\beta{\rightarrow}\infty$, and substituting for $B$ using \slBdef,
yields $\Bgen{\aa}(\beta)=2\beta^2/(\beta^2\!+%
\!4\pi{2n-1\over 2n})\,$.
Comparing with the data in table~1 then suggests the following
modification to \slBdef:
\eqn\atnewB{\hbox{
$\Bgen{\gk}(\beta)=2\beta^2/(\beta^2\!+\!4\pi{h\over \hv})$}\,.}
This does indeed turn out to be the case, and will be discussed
further below.

To continue, it is convenient to set
\eqn\isbldef{\bl{x}={\sinh\(\frac{\theta}/2+\frac{i\pi x}/{2}\)
\over\sinh\(\frac\theta/2-\frac i\pi x/{2}\)}~;}
the same as $\usbl{x}$, but without the $\hk$.
Assuming duality, {\it two} $k^{th}$ Coxeter numbers will be
relevant to each
non simply-laced theory, $\hk$ and $\hkv$. At weak coupling,
pole positions should approach integer multiples of $i\pi/\hk\,$; at
strong coupling (weak dual coupling), integer multiples of
$i\pi/\hkv\,.$ It is therefore
natural to define an interpolating block $\bl{x,y}\,,$
to replace $\usbl{x}\,$, now depending on two integers rather than just
one:
\eqn\iibldef{\bl{x,y}=\bl{(2{-}B)x/2\hk+By/2\hkv}}
The first index will `see' the classical data of $\gk$, the second
that of $\gk{}^{\vee}$. Now a generalisation of
\slbldef a\ can be given:
\eqn\bldef{\ubl{x,y}={\bl{x-1,y-1}\bl{x+1,y+1}\over
\bl{x-1,y+1}\bl{x+1,y-1}}~.}
Even for $\gk$ simply-laced, this provides quite a succinct way to
write the basic block, the correspondence being $\ubl{x}=\ubl{x,x}$
for these (self-dual) cases. However, one feature of
the non simply-laced S-matrices has yet to be captured:
`extra' cancellations between physical-strip poles and zeroes.
These never happen in the simply-laced cases -- the poles and zeroes
go through the bootstrap independently, allowing
minimal parts of these S-matrices to be defined -- but is crucial
for $\gk$ non simply-laced (excepting the self-dual $\aae$ theory, for
which an -- albeit not one-particle unitary -- minimal S-matrix can
indeed be defined\thinspace\RF\rFKMa\FKMa).
The extra freedom given by the {\it pair} of indices
entering \bldef\ permits the product
of two blocks to have fewer zeroes and poles than
expected, owing to cross-cancellations. There are precisely
two ways in which a product of two blocks \bldef\
can have a such a `partial' cancellation, one involving shifts of
the left index, and one shifts of the right. This motivates
the definitions
\eqn\fusdefs{\eqalign{%
\ibl {}{x,y}2 &=\ubl{x,y{-}1}\ubl{x,y{+}1}\cr
\ibl 2{x,y}{} &=\ubl{x{-}1,y}\ubl{x{+}1,y}\cr
}}
Each has the same number of poles and zeroes as $\ubl{x}$, owing to
cancellations which are easily spotted by using the definition \bldef.
Going further, products of more blocks can also have cancellations; for example
\eqn\morfus{\eqalign{\ibl {}{x,y}3 &=\ubl{x,y{-}2}\ubl{x,y}\ubl{x,y{+}2}\cr
\ibl 2{x,y}2 &=\ibl 2{x,y{-}1}\,\ibl 2{x,y{+}1}{}
 =\ibl {}{x{-}1,y}2\ibl {}{x{+}1,y}2\,.\cr}}
The most general object that
can be constructed in this way, with no more poles or zeroes than the
basic block, is, after cancellations,
\eqn\genbldef{\ibl a{x,y}b=
{\bl{x-a,y-b}\bl{x+a,y+b}\over
\bl{x-a,y+b}\bl{x+a,y-b}}~.}
(It is convenient to omit any subscripts equal to 1.)
All of the S-matrix elements listed in~\rDGZa\
can be rewritten, either as a product of the
`elementary' blocks $\ubl{x,y}$ alone, or, after fusings, in terms of
the more elaborate objects \genbldef. The resulting notation may be
viewed as a further refinement of the
`single index' scheme used in \rCDSa: the correspondence between
the two involves sending blocks of the form
$\ibl{}{{,}}b$ above into the blocks $\ibl{}{}{(b-1)/2}$ of \rCDSa\
in all cases but
$\bb/\aa$, where they become $\ibl{}{}{(b-2)/4}$ instead.

To see how the duality works, note from \atnewB\ that
$\Bgen{\gk}(4\pi/\beta)=2-\Bgen{\gk{}^{\vee}}(\beta)$, and so from
\iibldef\ it follows that expressions for the dual theory are
obtained just by swapping over the integers $x$ and $y$ in each block,
along with any subscripts present -- that is, read the formulae from
right to left instead of left to right.
Some `universal' features of the pole residues can also be derived.
Consider a single block $\ibl a{x,y}b\,$, in the
weak-coupling limit. This has two poles, which approach
$i\pi(x\!\pm\!a)/\hk$ as $\beta\!\rightarrow\!0\,.$
To leading order in $\beta$ (and with
$B(\beta)=\kappa\beta^2/2\pi+\dots\,$)
their residues are
\eqn\poleres{{\cal R}_{\pm}\(\ibl a{x,y}b\)
 =\pm {i\beta^2\kappa C\over 2\hkv}~,\quad{\rm with}~C=b~.}
(This is a little delicate, as
the position of the pole may also depend on $\beta\,$; the residue
given here corresponds to a simultaneous expansion, near the pole, in
$\beta$ and $\theta{-}\theta_0$, with $\theta_0$ the (mobile) pole
position. Following \rDGZa, this seems to be the appropriate
prescription for comparison with perturbation theory based on the
renormalised masses.) The new feature compared to the
simply-laced cases is the correction factor $C$ -- the `fusing'
of blocks changes their residues. There is another mechanism by which a
simple pole residue can be modified: other poles and zeroes, from other
blocks, which also approach the location of the pole in question as
$\beta\!\rightarrow\!0$. While these do not change the order of the
pole (for $\beta\!\ne\!0$, they are not in the same place), they do
change its residue, even to leading order in $\beta$. The general
formula for this change is a little unwieldy; one
example should suffice. Consider the product
$\ibl{}{x,y}b\ibl{}{x{+}2,y{+}p}{b'}\,$.
For small $\beta$, there are two simple poles near to
$i\pi(x{+}1)/\hk\,,$ one from each block. Without the presence of the
other block, their residues would be given by \poleres\ with $C=b,b'$.
Including this extra effect then multiplies $C$ by the
additional factors of
\eqn\polerescorr{{p\!+\!b'\!-\!b\over p\!-\!b'\!-\!b}\quad,\quad
                 {p\!+\!b\!-\!b'\over p\!-\!b\!-\!b'}}
respectively.
These two new phenomena -- the fusing and the interference of blocks
-- mesh with
another extra feature of the non simply-laced theories: the
already-mentioned occurrence of varying $\lambda^{abc}$ factors
correcting the area law \area.
This is illustrated in the next section by means of the
$\ff/\ee$ S-matrix.

\newsec{An example: the S-matrix for $\ff$ and $\ee$}
This S-matrix has been presented in \rCDSa, but will
be developed from scratch here both to demonstrate the approach just
outlined, and to highlight aspects of the comparison with perturbation
theory.  To start with, the
classical Lagrangian will be taken to be that of the
$\ff$ theory; however, an {\it a priori} assumption of duality allows
data from both algebras of the pair to
be used as input.  The two classical theories both
involve 4 non-degenerate masses, $m_1\dots m_4$. (The $\ee$ masses
are, via the folding idea\ts\NRF\rOTa\OTa\refs{\rOTa,\rBCDSc}, found
as the masses of particles 2,4,5 and 7 of the $e_7^{(1)}$ model.)
The full sets of couplings can be found in \rBCDSc, but as
an empirical rule the only `good' three-point couplings,
that survive to have bootstrap implications in the quantum theory, are
those which are found in {\it both} classical theories of a dual
pair. In terms of the generalised bootstrap principle put forward in
\rCDSa, this means that the corresponding S-matrix poles will turn out to
be `positive definite', their residues being positive multiples of $i$
thoughout the entire range of $\beta$.
For the case in hand, this selects the couplings $C^{111}$,
$C^{222}$, $C^{333}$, $C^{444}$, $C^{112}$, $C^{113}$, $C^{224}$,
$C^{123}$, and $C^{134}$. (It is interesting to note that, apart from
the four $\phi^3$-type couplings, the couplings listed are
those having depth\ts\rBCDSe\ one in the $e_7^{(1)}$ theory.) Their
classical fusing angles are:
\def\qsem{~;\quad}
\def\qqsem{\qquad\qquad}
\eqn\angles{\eqalign{U^1_{11}&=U^2_{22}=U^3_{33}=U^4_{44}=(8,12)\qqsem
U^2_{11}=(6,8)\qsem U^1_{12}=(9,14)\cr
U^3_{11}&=(2,2)\qsem~~~ U^1_{13}=(11,17)\qqsem~%
U^4_{22}=(2,4)\qsem U^2_{24}=(11,16)\cr
U^1_{23}&=(10,15)\qsem U^2_{31}=(9,13)\qsem~ U^3_{12}=(5,8)\cr
U^1_{34}&=(11,17)\qsem U^3_{41}=(10,16)\qsem U^4_{13}=(3,3)\cr}}
In this list, the first number of each pair, multiplied by $\pi/12$,
is the fusing angle for
$\ff$ (that is, the angle that emerges from \fusang\ when the
classical $\ff$ masses are used for $m_1\dots m_4\,$),
while the second, multiplied by $\pi/18$, is that for $\ee$ ($12$
and $18$ are the $k^{th}$ Coxeter numbers for $\ff$ and $\ee$).
These two fusing angles will be written as $U^a_{bc}(0)$ and $U^a_{bc}(2)$
respectively. If it is assumed that for the $\ff/\ee$ theory
there is again a linearising $B$-function $\Bgen{\ff}\!\equiv\!B$,
then the general fusing angle must be \vskip -4pt
\eqn\genfus{\hbox{$U^a_{bc}(B)={2{-}B\over
 2}U^a_{bc}(0)+{B\over 2}U^a_{bc}(2)\,.$}}
This is enough to postulate the ratios of the conserved
charges, via the bootstrap relations. For example,
using the couplings $C^{123}$ and $C^{224}$ leads to the quadruplet
$(\sin 2\theta_s,\sin(3{+}B/6)\theta_s,
\sin(7{-}B/6)\theta_s,\sin 2\theta_s\!+\!\sin(4{+}B/3)\theta_s)\,,$
where $\theta_s{=}\pi s/12$.
The spin $s$ runs over all integers coprime to $6$, the set of
exponents of $\ff$.  At $B\!=\!0\,,$ these are the `classical' conserved
charges of $\ff$, forming eigenvectors of the Cartan matrix of
the non-affine algebra.
Next, combining the implications of \angles\ and \genfus\ for pole
structure with the hypothesis that the S-matrix can
be written in terms of the blocks $\ubl{x,y}$ introduced in the last
section selects $S_{11}$ uniquely:
$S_{11}=\ubl{1,1}\ubl{5,7}\ubl{7,11}\ubl{11,17}$. The bootstrap can now
be followed through, a process which is simplified by the linear
notation. Note first that if $U^c_{ab}$ is specified by the pair of
integers $(m,n)$ in \angles, then $\bar U^c_{ab}=\pi{-}U^c_{ab}$ is
$(\bar m,\bar n)$ with $\bar m=12{-}m$, $\bar n=18{-}n$. Now to find
$S_{db}(\theta+i\bar U^c_{ab})$, say, first dismember the constituent
blocks
$\ubl{x,y}$ as $\ubl{x,y}_+/\ubl{-x,-y}_+$, where $\ubl{x,y}_+$ is defined
as in the earlier sequence of definitions \isbldef\ -- \bldef, but now
starting from $\bl{x}_+=\sinh(\frac\theta/2+\frac i\pi x/2)$. Such
blocks preserve their forms under shifts in $\theta$
(cf.~the discussion in \rDc\thinspace ), and for any $B$,
\eqn\bootrule{\ubl{x,y}_+(\theta{+}i\bar U^c_{ab}(B))\equiv\({\cal
T}_{\bar m,\bar n}\ubl{x,y}_+\)(\theta)
=\ubl{x{+}\bar m,y{+}\bar n}_+(\theta)~,}
owing to the linear natures of \genfus\ and \bldef. The shift
operator ${\cal T}_{\bar m,\bar n}$ thus defined manages to hide away
all $\theta$- and $B$- dependence, making for more elegant
calculations. With such tools to hand, the complete S-matrix
emerges as follows:
\def\ubbl#1{\ibl{}{#1}2}
\def\ubbll#1{\ibl 2{#1}2}
\eqnn\fullS
$$\eqalignno{
S_{11}&=\ubl{1,1}\ubl{5,7}\tc\cr
S_{12}&=\ubbl{4,6}\tc\cr
S_{13}&=\bigl(\ubl{2,2}\ubl{4,6}\tc\bigr)\ubbl{6,9}\cr
S_{14}&=\ubbl{3,4}\ubbl{5,8}\tc\cr
S_{22}&=\ubbl{1,2}\ubbl{5,8}\tc\cr
S_{23}&=\ubbl{3,5}\ubbl{5,7}\tc\cr
S_{24}&=\ubbl{2,4}\ubbl{4,6}\ubbl{6,8}\tc\cr
S_{33}&=\ubl{1,1}\ubbl{3,4}\ubbl{5,8}\ubl{5,7}\tc\cr
S_{34}&=\ubbl{2,3}\ubbl{4,5}\ubbl{4,7}\ubbl{6,9}\tc\cr
S_{44}&=\ubbl{1,2}\ubbl{3,4}(\ubbll{4,6})(\ubbl{5,8})^2\tc&\fullS\cr}$$
All the S-matrix elements are crossing-symmetric: the omitted blocks
can be restored using the general relation
$\ibl a{x,y}b(i\pi{-}\theta)=\ibl a{\hk{-}x,\hkv{-}y}b(\theta)$.

The dual form of this S-matrix, describing the situation for which the
classical theory is based on $\ee$, can be
found by reading \fullS\ backwards: for example,
$S_{12}=\ibl 2{6,4}{}\tc$, now with $\hk{=}18$
and $\hkv{=}12$. The dual conserved charges
emerge on substituting
$2\!-\!B$ for $B$ in the previous expressions: this yields
$(\sin 3\theta_s,\sin(5{-}B/4)\theta_s,
\sin(10{-}B/4)\theta_s,\sin 3\theta_s\!+\!\sin(7{-}B/4)\theta_s)\,,$
with $\theta_s\!=\!\pi s/18$. The spin $s$ takes the same values
as before, the exponents of $\ee$ being equal to those of $\ff$, but
setting $B\!=\!0$ now reveals the classical $\ee$
charges, the subset $(q^{(s)}_2,q^{(s)}_4,q^{(s)}_5,q^{(s)}_7)$
from the $e^{(1)}_7$ theory. While $e^{(1)}_7$ has extra exponents
over $\ee$, at integers equal to $9$ modulo $18$, the
corresponding conserved charges are identically zero for the particles
$2$, $4$, $5$ and $7$ which survive the fold to $\ee$. This can
be understood in the spirit of \rDc\
via the alternative characterisation of twisted Coxeter
elements described in \rDh: within the $E_7$ root
system, those roots annihilated by projection onto the spin-9
eigenplane of a Coxeter element $w$ -- the $w$-orbits for particles
$2$, $4$, $5$ and $7$ -- form a root system for $E_6$,
within which $w$ acts as a twisted Coxeter element of $E_6$.

There remains the question of the form of $B(\beta)$.
This can be approached via the
simple poles corresponding to the `good' couplings.
So long as attention is restricted to these poles,
the situation is only a little
more complicated than in the simply-laced cases.
To leading order in a perturbative treatment based on the
renormalised masses, the predicted residue of the simple pole in the
$a~b$ scattering amplitude due to a $\bar c$ bound state is
$i(C^{abc})^2/8m_a^2m_b^2\sin^2U^c_{ab}$. Using \area, this becomes
\eqn\pertres{{\cal R}={i(\lambda^{abc})^2\beta^2\over 2\hk}~.}
Consistency with \poleres\ then demands
$B(\beta)=\kappa\beta^2/2\pi{+}\dots\,$, with
\eqn\bform{\kappa={(\lambda^{abc})^2\over C}{\hkv\over\hk}~.}
To fix the numbers $\lambda^{abc}$,
the normalisation convention for the roots $\a_0,\dots\a_n$ must be
decided. That which leads to the
rule \atnewB\ for general $\gk$ starts by imposing $|\a_L|^2\!=\!2$ in all
untwisted ($k\!=\!1$) cases, where $\a_L$ is a longest root. Via
$\a_i^{\vee}\!=\!{2\over\a_i^2}\a_i\,$, this fixes $|\a_L|^2\!=\!2k$
for all but $\aae$; in this case, requiring that the roots are mapped
into themselves under duality
gives the same rule again.  This change of normalisations
over that used in \rBCDSc\ leads to the basic relation
\eqn\lambdarule{\lambda^{abc}=\sqrt{k}\,.}
Modifications are found in the untwisted non
simply-laced theories. Via the eigenvector property of the masses,
their particles are associated with the spots on the
non-affine Dynkin diagram of the corresponding finite algebra. They
can thus be called `short' or `long', depending on whether the
corresponding (non-affine) root is short or long. The change to
\lambdarule\ occurs when $a$, $b$ and $c$ are all short, and is
\refs{\rCMb,\rBCDSc,\rFb}
\eqn\lambdmod{\hbox{$a$, $b$, $c$ all short:}\quad\lambda^{abc}
 =\cases{{1\over\surd 2}& for $c^{(1)}_n$,
$f^{(1)}_4$;\cr
{2\over\surd 3}& for $g^{(1)}_2$.\cr}}
The other couplings continue to obey \lambdarule.
(For $b^{(1)}_n$ there are no short-short-short couplings at all,
and so \lambdmod\ is not needed.) The normalisations leading to
\lambdarule\ and \lambdmod\ only disagree
with \rDGZa\ for the $c_n^{(1)}/\dd$ pair, where
$|\a_L|^2{=}4$ was used for $c_n^{(1)}$, and $|\a_L|^2{=}2$ for $\dd$.

Of the four particle types in the $\ff$ theory, $1$ and $3$ are short.
Consider first $S_{11}(\theta)$. In the notation of \angles,
this has forward-channel poles at $(2,2)$, $(6,8)$ and $(8,12)$,
associated with particles $3$, $2$ and $1$ respectively. The S-matrix
predicts residues given by \poleres\ with $C=1,2,1$, the $2$ for the
$(6,8)$ pole coming via the mechanism \polerescorr. On the other
hand, \lambdmod\ gives the $\lambda$ factors for the $113$, $112$ and
$111$ couplings to be $1/\sqrt{2}$,
$1$, and $1/\sqrt{2}$ respectively. Thus in all three cases, \bform\
predicts $\kappa={1\over 2}\hkv/\hk=9/12=\hv/h$. The other mechanism by
which simple S-matrix
residues change, namely the fusing of blocks, comes into play
when $S_{12}$ is examined. This has forward-channel poles at $(5,8)$
and $(9,14)$, due to particles $3$ and $1$.
Whilst $1$ is short, $2$ is long and so the
$\lambda$ factors are $1$ for both the
$123$ and $121$ couplings. However,
the relevant blocks enter \poleres\ with
$b{=}2\,,$ so $C{=}2$ in
\bform, and $\kappa{=}9/12$ is again confirmed. The rest of \fullS\
can be checked in a similar fashion, and in all cases the combination
of \poleres, \polerescorr\ and \lambdmod\ conspires to produce
$(\lambda^{abc})^2/C=\frac 1/2$, exactly as required in \bform\ to
convert $\hkv/\hk$ into $\hv/h$.

On the other hand, the dual form of \fullS\ can be checked against the
classical $\ee$ data. This is even more straightforward, at least if
attention is again restricted to the `good' three-point couplings:
read right-to-left, all the relevant blocks enter \poleres\ with $b{=}1$,
and never feel the effect \polerescorr\ of neighbouring blocks, so $C{=}1$.
Furthermore, the $\lambda$ factors
behave in a uniform way for twisted algebras, being equal in this case
to $\sqrt{2}$. Thus $(\lambda^{abc})^2/C=2$ throughout, and
equation \bform\ yields $\kappa=2\times 12/18=12/9,$ in line with the
$\ee$ entry of table~1.

\newsec{Conclusions}
The analysis of the last section can be repeated for the block forms of
all the other non simply-laced Toda S-matrices,
and at every `good' simple pole,  the
mechanisms listed above -- the fusing of blocks, the influence of
neighbouring blocks, and the corrections to the basic area law --
combine to give the linearising $B$-function as defined earlier
the leading behaviour
$\Bgk(\beta)={\hv\over 2\pi h}\beta^2+\dots\,.$  The expression
\atnewB\ is then the natural extension of this with the desired
duality properties. Of course, for all the other cases the
full $\beta$-dependence has already given in~\rDGZa\
(for $g^{(1)}_2/d^{(3)}_4$, see also \rWWa), and one can
also check that the manipulations which lead to \atnewBa\ in the $\aa$
case also manage to confirm \atnewB\ for the others. However, in doing
this the already-noted change in root normalisations for
$\cc/\dd$ should be taken into account.

The main object of this note, apart from establishing the
$\beta$-dependence of the $\ff/\ee$ S-matrix, has been
to point out that the non simply-laced S-matrices share rather more
of the universal features of the simply-laced cases than might have
been thought. This reinforces the idea that there may ultimately be
some geometrical structure underlying these theories, as described for
the simply-laced cases in \refs{\rDc,\rDh}. It is hoped that the alternative
ways of viewing the S-matrices outlined here will lead to some
insight into this question.

\bigskip\goodbreak\noindent
{\bf Acknowledgements}\nobreak\medskip\nobreak
I would like to thank G\'erard Watts, Robert Weston, Gustav Delius
and Marc Grisaru for discussions, and Ed Corrigan and Ryu Sasaki both
for discussions and for helpful comments on the manuscript. This work was
supported by a grant under the Science Programme of the European
Community.

\listrefs
\bye